\documentclass[11pt]{article}
\usepackage[left=1in,top=1in,right=1in,bottom=1in]{geometry}
\usepackage{times}

\usepackage{amsmath}
\usepackage{amssymb}
\usepackage{amsthm}
\usepackage{graphicx}
\usepackage{verbatim}
\usepackage{url}
\usepackage{algorithm}
\usepackage[noend]{algpseudocode}
\allowdisplaybreaks

\usepackage{siunitx}

\allowdisplaybreaks

\begin{document}

\title{Quadratic Programming Approach for Nash Equilibrium Computation in Multiplayer Imperfect-Information Games} 

\author{Sam Ganzfried\\
Ganzfried Research \\
sam.ganzfried@gmail.com
}

\date{\vspace{-5ex}}

\maketitle

\begin{abstract}
There has been significant recent progress in algorithms for approximation of Nash equilibrium in large two-player zero-sum imperfect-information games
and exact computation of Nash equilibrium in multiplayer strategic-form games. While counterfactual regret minimization and fictitious play are scalable to large
games and have convergence guarantees in two-player zero-sum games, they do not guarantee convergence to Nash equilibrium in multiplayer games.
We present an approach for exact computation of Nash equilibrium in multiplayer imperfect-information games that solves a quadratically-constrained program based on a nonlinear
complementarity problem formulation from the sequence-form game representation. This approach capitalizes on recent advances for solving nonconvex quadratic programs. Our algorithm is able to
quickly solve three-player Kuhn poker after removal of dominated actions. Of the available algorithms in the Gambit software suite, only the logit quantal response approach 
is successfully able to solve the game; however, the approach takes longer than our algorithm and also involves a degree of approximation. Our formulation
also leads to a new approach for computing Nash equilibrium in multiplayer strategic-form games which we demonstrate to outperform a previous quadratically-constrained program formulation.
\end{abstract}

\section{Introduction}
\label{se:intro}
Nash equilibrium is the central solution concept in game theory. In recent years there has been substantial progress on algorithms for its computation and approximation in various classes of games. Two-player zero-sum games are a special case due to applicability of the minimax theorem and the fact that Nash equilibrium can be computed in polynomial time. For both two-player zero-sum strategic-form and extensive-form games with perfect recall, there exists a linear programming formulation which can be solved efficiently to compute a Nash equilibrium~\cite{Koller94:Fast}. For large two-player zero-sum extensive-form games Nash equilibrium can be efficiently approximated by procedures such as counterfactual regret minimization~\cite{Zinkevich07:Regret,Lanctot09:Monte} and fictitious play~\cite{Brown51:Iterative,Robinson51:Iterative}. For two-player non-zero-sum strategic-form games, there are three exact algorithms with selective superiority: the Lemke-Howson algorithm~\cite{Lemke64:Equilibrium}, which is based on solving a linear complementarity problem, a support-enumeration approach~\cite{Porter08:Simple}, and a mixed-integer program formulation~\cite{Sandholm05:Mixed}. Two-player non-zero-sum imperfect-information games can be solved by Lemke's algorithm, which is also based on solving a linear complementarity problem~\cite{Lemke65:Bimatrix,Koller94:Fast}. In the last decade there has been a surge of progress on algorithms for exact computation of Nash equilibrium in multiplayer strategic-form games. The exclusion method~\cite{Berg17:Exclusion} was demonstrated to outperform existing algorithms in the Gambit software suite~\cite{Savani25b:Gambit} on randomly-generated games and games produced by the GAMUT generator~\cite{Nudelman04:Run}. This approach has been subsequently improved upon by approaches based on solving bilinear program formulations~\cite{Ganzfried24:Fast,Zhang23:Computing,Cerny25:Spatial}.

In this paper we focus on the problem of computing exact Nash equilibrium in multiplayer imperfect-information games. These games exhibit several degrees of complexity: computing a Nash equilibrium is PPAD-hard~\cite{Chen05:Nash,Chen06:Settling,Daskalakis09:Complexity}, and solving imperfect-information games is significantly more challenging than strategic-form games. Iterative self-play algorithms such as fictitious play and counterfactual regret minimization can still be run on these games; however they have no guarantees of convergence to Nash equilibrium, and there are small 3-player games for which they are known to not solve. These algorithms scale efficiently to large problem instances and are still likely good choices in practice for large problems; however, they may not be effective for smaller problems if an exact Nash equilibrium is desired. The Gambit software suite offers several algorithms that apply to multiplayer imperfect-information games; however, several of them involve first converting the game to strategic form, which involves an exponential blowup and is only feasible for tiny games. For the problem ``Compute one Nash equilibrium'' Gambit offers three methods: simplicial subdivision, tracing logit equilibria, and solving systems of polynomial equations. Of these approaches, only the final two include capability of solving the extensive-form game, while simplicial subdivision~\cite{Laan87:Simplicial} requires converting the game to strategic form. Gambit also includes functionality to ``Compute some Nash equilibria.'' The methods include: looking for pure strategy equilibria, minimizing the Lyapunov function, solving systems of polynomial equations, global Newton tracing, and iterated polymatrix approximation. Of these, only the first three include capability of solving the extensive-form game, while the final two require converting the game to strategic form. We note that it is stated~\cite{Blum06:Continuation} that the global Newton~\cite{Govindan02:Structure,Govindan03:Global} and iterated polymatrix approximation~\cite{Govindan04:Computing} methods are also applicable to extensive-form games; but this functionality does not appear to be supported by Gambit. In our experiments we will examine performance of all of Gambit's applicable methods. As we will see, the only algorithm capable of solving 3-player Kuhn poker after removal of dominated actions is the logit tracing procedure for computing quantal response equilibrium~\cite{Turocy10:Computing}. Logit quantal response equilibrium (QRE) is a distinct solution concept from Nash equilibrium; however, a logit QRE is parameterized by a scalar precision parameter $\lambda$, and the limit of any branch in the QRE correspondence approaches a Nash equilibrium as $\lambda$ approaches infinity. Theoretically this approach has a global convergence guarantee to a Nash equilibrium (in fact to a sequential equilibrium which is a refinement of Nash equilibrium); in practice Gambit halts the algorithm when a regret threshold is reached.

Our approach involves formulating the problem as a nonlinear complementarity problem (NCP) based on a generalization of the sequence-form game representation~\cite{Koller94:Fast} to multiple players. We then model this NCP as a quadratically-constrained feasibility program, which we can solve using recent advances in algorithms for nonconvex quadratic optimization~\cite{Gurobi25:Gurobi}. Note that we are certainly not the first to formulate the problem of Nash equilibrium computation as an NCP. McKelvey and McLennan~\cite{McKelvey96:Computation} discuss numerous prior approaches (several of which are supported by Gambit) for computing a Nash equilibrium in multiplayer strategic-form games by solving a corresponding NCP. However, they note: ``These methods do not satisfy global convergence.'' By contrast, our approach is an exact approach that guarantees computation of Nash equilibrium (subject to numerical precision). The prior approaches fall either into the category of heuristics with no convergence guarantee (e.g., (counterfactual) regret minimization, fictitious play, global Newton method), or an approximation algorithm that guarantees convergence in the limit (e.g., logit QRE). 

Our primary experiments will be on 3-player Kuhn poker, a nontrivial simplified version of 3-player poker that has been used as a testbed game in the AAAI Annual Computer Poker Competition for several years. Specifically, we will experiment on the reduced version of the game after dominated actions have been removed. We note that an infinite family of Nash equilibria has been computed analytically for the full game~\cite{Szafron13:Parametrized}. The analysis to calculate these equilibria is quite involved, and clearly it is desirable to also develop general-purpose algorithms that can broadly be applied to multiplayer imperfect-information games. We also note that prior work has demonstrated that counterfactual regret minimization is able to compute a strategy profile very close to Nash equilibrium (no player could gain more than 0.0045 by deviating)~\cite{Abou10:Using}. So we are certainly not asserting that our algorithm is the first to solve this game. But we argue that it is clearly desirable to have a general exact algorithm that can solve this game quickly, since it is not feasible to perform detailed manual analysis in general and counterfactual regret minimization happened to converge to approximate Nash equilibrium in this game without any guarantee of convergence. We will show that our new approach solves this game in under 2.5 seconds (computes a strategy profile such that no player can gain more than $\num{1.4e-17}$ by deviating), while the only Gambit approach that is able to solve the game is logit QRE which takes over 40 seconds and computes a strategy profile where no player can gain more than $\num{1.67e-5}$ by deviating.

\section{Algorithm}
\label{se:algorithm}
Imperfect-information games are modeled using extensive-form game trees, where play proceeds from the root node to a terminal leaf node at which point all players receive payoffs. Each non-terminal node has an associated player (possibly \emph{chance}) that makes the decision at that node. These nodes are partitioned into \emph{information sets}, where the player whose turn it is to move cannot distinguish among the states in the same information set. Therefore, in any given information set, a player must choose actions with the same distribution at each state contained in the information set. If no player forgets information that they previously knew, we say that the game has \emph{perfect recall}. A (mixed) \emph{strategy} for player $i,$ $\sigma_i \in \Sigma_i,$ is a function that assigns a probability distribution over all actions at each information set belonging to $i$. 

In order to present our new approach we review the \emph{sequence-form representation} for two-player extensive-form games of imperfect information~\cite{Koller94:Fast}. Rather than operate on the full pure strategy space, which has size exponential in the size of the game tree, the sequence-form works with sequences of actions along trajectories from the root node to leaf nodes. For player 1, the matrix $\mathbf{E}$ is defined where each row corresponds to an information set (including an initial row for the ``empty'' information set), and each column corresponds to an action sequence (including an initial row for the ``empty'' action sequence). In the first row of $\mathbf{E}$ the first element is 1 and all other elements are 0; subsequent rows have -1 for the entries corresponding to the action sequence leading to the root of the information set, and 1 for all actions that can be taken at the information set (and 0 otherwise). Thus $\mathbf{E}$ has dimension $c_1 \times d_1$, where $c_i$ is the number of information sets for player $i$ and $d_i$ is the number of action sequences for player $i$. Matrix $\mathbf{F}$ is defined analogously for player 2. The vector $\mathbf{e}$ is defined to be a column vector of length $c_1$ with 1 in the first position and 0 in other entries, and vector $\mathbf{f}$ is defined with length $c_2$ analogously. The matrix $\mathbf{A}$ is defined with dimension $d_1 \times d_2$ where entry $A_{ij}$ gives the payoff for player 1 when player 1 plays action sequence $i$ and player 2 plays action sequence $j$ multiplied by the probabilities of chance moves along the path of play. The matrix $\mathbf{B}$ of player 2's payoffs is defined analogously. In zero-sum games $\mathbf{B} = -\mathbf{A}.$

Given these matrices we can solve one of two linear programming problems to compute a Nash equilibrium in zero-sum extensive-form games~\cite{Koller94:Fast}. In the first formulation the primal variables $\mathbf{x}$ correspond to player 1's mixed strategy while the dual variables correspond to player 2's strategy. In the second formulation, which is the dual problem of the first formulation, the primal decision variables $\mathbf{y}$ correspond to player 2's strategy while the dual variables correspond to player 1's strategy.

\[
\begin{array}{rrl} 
&\max_{\mathbf{x},\mathbf{q}}& -\mathbf{q}^T \mathbf{f} \\ 
&\mbox{s.t.}& \mathbf{x}^T (-\mathbf{A}) - \mathbf{q}^T \mathbf{F} \leq \mathbf{0} \\
& & \mathbf{x}^T \mathbf{E}^T = \mathbf{e}^T \\
& & \mathbf{x} \geq \mathbf{0}\\
\end{array} 
\]

\[
\begin{array}{rrl} 
&\min_{\mathbf{y},\mathbf{p}}& \mathbf{e}^T \mathbf{p} \\ 
&\mbox{s.t.}& -\mathbf{A} \mathbf{y} + \mathbf{E}^T \mathbf{p} \geq \mathbf{0} \\
& & -\mathbf{F} \mathbf{y} = -\mathbf{f} \\
& & \mathbf{y} \geq \mathbf{0}\\
\end{array} 
\]

For two-player non-zero-sum games, the problem of finding a Nash equilibrium is the feasibility problem of finding
$\mathbf{x},$ $\mathbf{y},$ $\mathbf{p},$ $\mathbf{q}$, such that~\cite{Koller94:Fast}:

\[
\begin{array}{rrl} 
-\mathbf{A} \mathbf{y} + \mathbf{E}^T \mathbf{p} & \geq & \mathbf{0} \\
-\mathbf{B}^T \mathbf{x} + \mathbf{F}^T \mathbf{q} & \geq &\mathbf{0} \\
-\mathbf{E} \mathbf{x} & = &-\mathbf{e} \\
-\mathbf{F} \mathbf{y} & = &-\mathbf{f} \\
\mathbf{x} & \geq &\mathbf{0}\\
\mathbf{y} &\geq &\mathbf{0}\\
\mathbf{x}^T (-\mathbf{A} \mathbf{y} + \mathbf{E}^T \mathbf{p}) & = &0\\
\mathbf{y}^T (-\mathbf{B} \mathbf{x} + \mathbf{F}^T \mathbf{q}) & = &0\\
\end{array} 
\]

The final two constraints are called complementarity slackness conditions (CSC), and the full system is known
as a linear complementarity problem (LCP). It is no longer a linear program since the CSCs involve products of variables.
This LCP can be solved using Lemke's algorithm~\cite{Lemke65:Bimatrix} or the related Lemke-Howson algorithm~\cite{Lemke64:Equilibrium}.
We would like to extend this result to develop a feasibility problem for games with $n > 2$ players. We will start with the case $n = 3.$ The sequence-form representation extends straightforwardly to 3 players. We define matrix $\mathbf{G}$ for player 3 analogously to $\mathbf{E}$ and $\mathbf{F},$ and define vector $\mathbf{g}$ analogously to $\mathbf{e},\mathbf{f}.$ The utility functions can no longer be represented as 2-dimensional matrices. We write $u_1(i,j,k)$ as player 1's utility when player 1 plays action sequence $i,$ player 2 plays action sequence $j$, and player 3 plays action sequence $k.$ We represent player 2 and 3's utilities analogously as $u_2(i,j,k), u_3(i,j,k).$

Consider the problem of player 1 playing a best response when player 2 plays $\mathbf{y}$ and player 3 plays $\mathbf{z}$:

\[
\begin{array}{rrl} 
&\max_{\mathbf{x}} &\sum_i \sum_j \sum_k x_i y_j z_k u_1(i,j,k)  \\ 
&\mbox{s.t.}& \mathbf{E} \mathbf{x} = \mathbf{e} \\
& & \mathbf{x} \geq \mathbf{0}\\
\end{array} 
\]

Let us rewrite this as a convex minimization problem by negating the objective:

\[
\begin{array}{rrl} 
&\min_{\mathbf{x}} &-\sum_i \sum_j \sum_k x_i y_j z_k u_1(i,j,k)  \\ 
&\mbox{s.t.}& \mathbf{E} \mathbf{x} = \mathbf{e} \\
& & \mathbf{x} \geq \mathbf{0}\\
\end{array} 
\]

The Lagrangian is
$$L(\mathbf{x},\boldsymbol\tau^1,\mathbf{r}^1) = -\sum_i \sum_j \sum_k x_i y_j z_k u_1(i,j,k) - (\mathbf{E} \mathbf{x} - \mathbf{e})^T \boldsymbol\tau^1
- (\mathbf{r}^1)^T \mathbf{x}$$

Defining $\boldsymbol\lambda^1 = -\boldsymbol\tau^1$:

$$L(\mathbf{x},\boldsymbol\lambda^1,\mathbf{r}^1) = -\sum_i \sum_j \sum_k x_i y_j z_k u_1(i,j,k) - (\mathbf{E} \mathbf{x} - \mathbf{e})^T (-\boldsymbol\lambda^1) - (\mathbf{r}^1)^T \mathbf{x}$$
$$= -\sum_i \sum_j \sum_k x_i y_j z_k u_1(i,j,k) + (\mathbf{E} \mathbf{x} - \mathbf{e})^T \boldsymbol\lambda^1 - (\mathbf{r}^1)^T \mathbf{x}$$

$$\frac{\partial L}{\partial x_i} = -\sum_j \sum_k y_j z_k u_1(i,j,k) + \sum_{j} \lambda^1_j E_{ji} - r^1_i$$
 
So the first-order necessary optimality conditions for player 1's best response problem are:

\[
\begin{array}{rrl} 
\mathbf{E} \mathbf{x} &= &\mathbf{e}\\
\mathbf{x} &\geq &\mathbf{0}\\
\mathbf{r}^1 &\geq &\mathbf{0}\\
-\sum_j \sum_k y_j z_k u_1(i,j,k) + \sum_{j} \lambda^1_j E_{ji} - r^1_i & = &0 \mbox{ for all } i \\
x_i r^1_i &= &0 \mbox{ for all } i\\
\end{array} 
\]

Since the problem is a convex minimization problem these conditions are sufficient for optimality as well.
Adding in analogous constraints for players 2 and 3, the problem is to find $\mathbf{x},\mathbf{y},\mathbf{z}$,
$\boldsymbol\lambda^1,\boldsymbol\lambda^2,\boldsymbol\lambda^3$, $\mathbf{r}^1,\mathbf{r}^2,\mathbf{r}^3$ such that:

\[
\begin{array}{rrl} 
-\sum_j \sum_k y_j z_k u_1(i,j,k) + \sum_{j} \lambda^1_j E_{ji} - r^1_i & = &0 \mbox{ for all } i \\
-\sum_i \sum_k x_i z_k u_2(i,j,k) + \sum_{i} \lambda^2_i F_{ij} - r^2_j & = &0 \mbox{ for all } j \\
-\sum_i \sum_j x_i y_j u_3(i,j,k) + \sum_{j} \lambda^3_j G_{jk} - r^3_k & = &0 \mbox{ for all } k \\
\mathbf{E} \mathbf{x} &= &\mathbf{e}\\
\mathbf{F} \mathbf{y} &= &\mathbf{f}\\
\mathbf{G} \mathbf{z} &= &\mathbf{g}\\
\mathbf{x} &\geq &\mathbf{0}\\
\mathbf{y} &\geq &\mathbf{0}\\
\mathbf{z} &\geq &\mathbf{0}\\
\mathbf{r}^1 &\geq &\mathbf{0}\\
\mathbf{r}^2 &\geq &\mathbf{0}\\
\mathbf{r}^3 &\geq &\mathbf{0}\\
x_i r^1_i &= &0 \mbox{ for all } i\\
y_i r^2_i &= &0 \mbox{ for all } i\\
z_i r^3_i &= &0 \mbox{ for all } i\\
\end{array} 
\]

This feasibility program is not a linear program due to several products of variables:
$y_j z_k,$ $x_i z_k$, $x_i y_j$, $x_i r^1_i,$ $y_i r^2_i,$ $z_i r^3_i.$ 
This formulation can be straightforwardly generalized to $n > 3$ players by adding in the original decision constraints,
multiplier sign conditions, Lagrange derivative conditions, and complementary slackness conditions for each player.
The Lagrange derivative conditions will involve products of $n-1$ variables, while the complementary slackness conditions 
still involve products of 2 variables. However, we can still model the Lagrange derivative conditions with quadratic constraints by incrementally defining new product variables. E.g., define $w_{ij} = x_i y_j,$ then $v_{ijk} = w_{ij} z_k$, etc. So for arbitrary $n > 2$, we can define the problem of finding a Nash equilibrium as a nonlinear complementarity problem, which can be modeled as a quadratically-constrained feasibility program. For efficiency we prefer to introduce as few new auxiliary bilinear variables as possible. For example, for four players with variables $\mathbf{x}^1,\mathbf{x}^2,\mathbf{x}^3,\mathbf{x}^4,$ we can create a new bilinear program with only the introduction of new variables $y_{ij} = x^1_i x^2_j$ and $z_{ij} = x^3_i x^4_j.$ We have $x^1_i x^2_j x^3_k = y_{ij} x^3_k,$ $x^1_i x^2_j x^4_m = y_{ij} x^4_m,$ $x^1_i x^3_k x^4_m = x^1_i z_{km},$ $x^2_j x^3_k x^4_m = x^2_j z_{km}.$ 

This nonlinear complementarity program (NCP) also applies to strategic-form games if we interpret the strategies as mixed strategies over the probability simplex as in~\cite{Koller94:Fast}. For this formulation, $\mathbf{x},\mathbf{y},\mathbf{z}$ are now probability vectors, $\mathbf{E},\mathbf{F},\mathbf{G}$ are one-dimensional vectors of all ones, $\mathbf{e},\mathbf{f},\mathbf{g}$ are the scalar 1, $\boldsymbol\lambda^1,\boldsymbol\lambda^2,\boldsymbol\lambda^3$ are scalars, and $\mathbf{r}^1,\mathbf{r}^2,\mathbf{r}^3$ are scalars. With these interpretations, we can view the program as a quadratically-constrained feasibility program for multiplayer Nash equilibrium computation in strategic-form games.

Recent advances in algorithms for solving nonconvex quadratic programs can enable us to directly solve these formulations~\cite{Gurobi25:Gurobi}.

\section{Three-player Kuhn poker}
\label{se:kuhn}
Three-player Kuhn poker is a simplified form of limit poker that has been used as a testbed game in the AAAI Annual Computer Poker Competition for several years. There is a single round of betting. Each player first antes a single chip and is dealt a card from a four-card deck that contains one Jack (J), one Queen (Q), one King (K), and one Ace (A). The first player has the option to \emph{bet} a fixed amount of one additional chip (by contrast in \emph{no-limit} games players can bet arbitrary amounts of chips) or to \emph{check} (remain in the hand but not bet an additional chip). When facing a bet, a player can \emph{call} (i.e., match the bet) or \emph{fold} (forfeit the hand). No additional bets or raises beyond the additional bet are allowed (while they are allowed in other common poker variants such as Texas hold 'em). If all players but one have folded, then the player who has not folded wins the \emph{pot}, which consists of all chips in the middle. If more than one player has not folded by the end there is a \emph{showdown}, at which point the players reveal their private card and the player with the highest card wins the entire pot (which consists of the initial antes plus all additional bets and calls). The Ace is the highest card, followed by King, Queen, and Jack. As one example of a play of the game, suppose the players are dealt Queen, King, Ace respectively, and player 1 checks, player 2 checks, player 3 bets, player 1 folds, and player 2 calls; then player 3 would win a pot of 5, for a profit of 3 (while player 1 loses 1 and player 2 loses 2).

Note that despite the fact that 3-player Kuhn poker is only a synthetic simplified form of poker and is not actually played competitively, it is still far from trivial to analyze, and contains many of the interesting complexities of popular forms of poker such as Texas hold 'em. First, it is a game of imperfect information, as players are dealt a private card that the other agents do not have access to, which makes the game more complex than a game with perfect information that has the same number of nodes. Despite the size, it is not trivial to compute Nash equilibrium analytically, though recently an infinite family of Nash equilibria has been computed~\cite{Szafron13:Parametrized}. The equilibrium strategies exhibit the phenomena of \emph{bluffing} (i.e., betting with weak hands such as a Jack or Queen), and \emph{slow-playing} (aka \emph{trapping}) (i.e., checking with strong hands such as a King or Ace in order to induce a bet from a weaker hand). The family of equilibria is based on several parameter values, which once selected determine the probabilities for the other portions of the strategies. One can see that randomization and including some probability on trapping and bluffing are essential in order to have a strong and unpredictable strategy. Thus, while this game may appear quite simple at first glance, analysis is still very far from simple, and the game exhibits many of the complexities of far larger games that are played competitively by humans for large amounts of money. 

Prior work notes that several of the Nash equilibrium strategy probabilities must take on ``necessary parameter values'' of 0 or 1 (i.e., certain actions are \emph{dominated})~\cite{Szafron13:Parametrized}. These actions include calling bet with Jack, folding to a bet with Ace, calling a bet with Queen after a bet and a call, and checking with Ace after two players check. We can remove these clearly dominated actions from the game, constructing a reduced game that is guaranteed to contain a Nash equilibrium of the full game as well. Our experiments will focus on this reduced game with the dominated actions removed.

\section{Experiments}
\label{se:exp}
For all experiments with our algorithms we used the non-convex quadratic solver, which guarantees global optimality~\cite{Gurobi25:Gurobi}. We used an Intel Core i7-1065G7 processor with base clock speed of 1.30 GHz (and maximum turbo boost speed of up to 3.9 GHz) with 16 GB of RAM under 64-bit Windows 11 (4 cores/8 threads).

We will first briefly present results on strategic-form games before our main results on three-player Kuhn poker. For strategic-form games we compare our bilinear program derived from the nonlinear complementarity problem formulation (NCP) against a prior direct mixed-integer bilinear program approach (MIP)~\cite{Ganzfried24:Fast}. As prior experiments have done we set Gurobi's feasibility tolerance parameter to $\num{1e-4}$ for the MIP approach (while keeping it at its default value of $\num{1e-6}$ for the NCP approach). We generated games with payoffs uniformly random in [0,1] for a variety of number of players $n$ and pure strategies $m$, as prior work has done. For each set of parameter values $(n,m)$, we generated 1,000 random games and calculate the average run time of both approaches (with no time limit restriction). The results are given in Table~\ref{ta:results-sfg}. For all game classes the NCP approach had smaller runtimes than the MIP approach, with increased performance improvement for larger games. For $n = 3$, $m = 6,$ NCP runtimes were 2.7x faster, and for $n = 4,$ $m = 3$ nearly 5x. The results indicate that between these two direct optimization approaches NCP is preferable. This suggests that when considering optimization formulations for Nash equilibrium it is desirable to eliminate integral-constrained variables and use only continuous variables, which has also been observed by others~\cite{Cerny25:Spatial}. We note that as for the MIP approach NCP is also outperformed by newer more complex bilinear methods~\cite{Zhang23:Computing,Cerny25:Spatial}. We present these results not to suggest that this NCP approach is the best approach for solving multiplayer strategic-form games; only for a baseline comparison to a related recent direct bilinear optimization approach~\cite{Ganzfried24:Fast}.

\begin{table}[!ht]
\centering
\caption{Runtimes in seconds of bilinear program approaches for strategic-form games with all payoffs uniform in [0,1]. The first column ($n$) is the number of players and second
column ($m$) is the number of pure strategies per player.}
\label{ta:results-sfg}
\begin{tabular}{|*{4}{c|}} \hline
$n$ &$m$ &NCP avg. runtime &MIP avg. runtime\\ \hline
3 &3 &0.0619 &0.1002 \\ \hline
3 &4 &0.1864 &0.2487 \\ \hline
3 &5 &0.5215 &1.2641 \\ \hline
3 &6 &6.9851 &18.8500 \\ \hline
4 &3 &0.3254 &1.5648 \\ \hline
5 &2 &0.1858 &0.2983 \\ \hline
\end{tabular}
\end{table}

We now consider our main experiments with the imperfect-information version of our approach on three-player Kuhn poker. The full game has 48 total information sets (16 per player) and 601 total nodes, which includes 288 player decision nodes, 312 terminal nodes, and one chance node. The reduced game after removal of dominated actions has 48 total information sets and 415 total nodes (252 player decision nodes, 162 terminal nodes, and one chance node). So the total number of information sets remains the same while the number of player decision nodes is decreased by 12.5\%.

Our approach was able to solve the game in 2.47 seconds. We verified that no player can gain more than $\num{1.4e-17}$ by deviating from the computed solution, indicating that it is essentially an exact Nash equilibrium. The model input to Gurobi contained 72 rows and 249 columns with 168 nonzeros and 198 quadratic constraints. Gurobi's initial presolve method removed 45 rows and 57 columns. Gurobi then recognizes the model is non-convex and reformulates it as a MIP. A second presolve procedure removes 24 rows and 76 columns taking 0.00 seconds. This new model has 910 rows, and 375 columns, with 2306 nonzeros, 66 Special-Ordered Set (SOS) constraints, 201 bilinear constraints, 375 continuous variables, and 0 integer variables. Gurobi solved this model in 2.47 seconds (1.42 work units), exploring 4624 nodes in 207,149 simplex iterations (using 8 threads). 32 cutting planes were used: 7 Gomory, 7 flow cover, and 18 Relaxation Linearization Technique (RLT) cuts. 

By contrast, the approach was not able to solve the full game (without removal of dominated actions) in 24 hours. The model input to Gurobi contained 51 rows with 249 columns and 149 nonzeros. Note that we constructed the reduced game by simply constructing the full game and then adding equality constraints in the Gurobi model setting certain action sequence probabilities to 0. This explains why the reduced game model initially contains more rows than the full model. The initial presolve phase removed only 3 rows and 6 columns. After recognizing that the model is non-convex and deciding to solve it as a MIP, the presolve phase then adds in 12 rows and 0 columns, followed by removing 0 rows and 9 columns (taking 0.00 seconds). The final presolved model has 1728 rows and 637 columns with 4357 nonzeros, 96 SOS constraints, 396 bilinear constraints, 637 continuous variables, and 0 integer variables. 

These results indicate on the positive side that the algorithm is able solve the reduced game quite quickly; however, they also indicate its limited scalability by the inability to solve the full game. After the presolve procedure the full game model contains about twice as many nonzero entries as the reduced game, giving some indication of the limit on the size of games that can be solved by the approach. The results also indicate the potential power of removing dominated actions as a preprocessing step prior to Nash equilibrium computation in imperfect-information games~\cite{Ganzfried25:Dominated}. This procedure makes the difference between our approach being able to quickly solve the game and not solve it at all in this instance.

The solution of our approach in the reduced game gave expected payoff to players 1 and 2 of $-0.0208333$ $\approx -\frac{1}{48}$ and to player 3 of $0.0416667$ $\approx \frac{1}{24}.$ We compared the solution to the infinite family of solutions computed analytically in prior work~\cite{Szafron13:Parametrized}. Our strategies agreed with the strategies in their family for all situations except two. For player 1, when holding King and facing check-bet-fold, our strategy always folds while their strategy calls and folds with probability $\frac{1}{2}$ (denoted by $a_{33} = \frac{1}{2}$ in their Table 3); and for player 3, when holding King and facing bet-fold, our strategy calls with probability $\frac{1}{2}$ while their strategy always folds ($c_{32} = 0$ in their Table 3). Despite these differences, our equilibrium payoffs agree with theirs for $\beta = \max \{b_{11},b_{21}\} = 0,$ where $b_{11}$ is the probability player 2 bets with Jack after a check and $b_{21}$ is probability player 2 bets with Queen after a check. The reason for this is that the information sets at which our strategy differs from the ones in their family are reached with probability zero under our strategy profile. Since player 2 never bets following a check, player 1 will never have a King facing check-bet-fold, and a range of strategies are possible that maintain equilibrium. Similarly, since player 1 will never bet initially, player 3 will never face bet-fold with a King. Since the only differences between the strategies are at information sets that are not reached, the Nash equilibrium expected payoffs will be identical. This result indicates that additional Nash equilibria exist beyond those in the infinite family previously computed (though perhaps they only differ from the family off the path of play). 

We next experimented on the reduced game using the algorithms from version 16.4 of the Gambit software suite~\cite{Savani25b:Gambit}, which implements several classic algorithms for two and multiplayer games in both strategic and extensive form. As stated in Section~\ref{se:intro}, some of these approaches apply only to two-player games, some apply only to strategic-form games, and some apply to extensive-form games by first converting the game to strategic form (which is not practical on reduced three-player Kuhn poker). There are options of running these algorithms on a command line or directly on Gambit's GUI. The advantage of the latter is that the resulting strategies are superimposed directly on the game tree in a human-readable way, while the output from the command line is a long list of numbers that is difficult to decipher. The GUI displays all probabilities to four significant digits, which may be indicative of approximation and/or roundoff errors. 

Most of Gambit's approaches were inapplicable or ineffective for solving reduced three-player Kuhn poker. The algorithms `enummixed,' `lcp,' and `lp' can only run on two-player games. We also note that the `enumpoly' algorithm which solves a system of polynomial equations is listed on the GUI and in the documentation, but does not appear to be available for implementation in Gambit 6.4 (attempting to run it caused the GUI to crash). For simplicial subdivision (`simpdiv'), global Newton method (`gnm'), and iterated polymatrix approximation (`ipa'), the GUI only gives the option to run them using the strategic form, not the extensive form. Running simplicial subdivision from the command line causes the program to crash, and we halted the other algorithms after ten minutes. The only two remaining algorithms that are applicable to multiplayer extensive-form games are Lyapunov function minimization (`liap') and quantal response equilibrium path following (`logit'). The Lyapunov minimization algorithm failed to find a solution after ten minutes on both the GUI and command line, so we halted it as well.

The only remaining method was the `logit' approach that computes the principal branch of the (logit) quantal response correspondence~\cite{Turocy10:Computing}. Logit quantal response equilibrium (QRE) is a distinct solution concept from Nash equilibrium; however, a logit QRE is parameterized by a scalar precision parameter $\lambda$, and the limit of any branch in the QRE correspondence approaches a Nash equilibrium as $\lambda$ approaches infinity. Theoretically this approach has a global convergence guarantee to a Nash equilibrium (in fact to a sequential equilibrium which is a refinement of Nash equilibrium); in practice Gambit halts the algorithm when a regret threshold is reached. This approach successfully computed a solution in 43 seconds using the GUI (and 49 seconds on the command line). The solution from the GUI was easy to comprehend and we were able to verify that no player can gain more than $\num{1.667e-5}$ by deviating. While this value is significantly larger than the $\epsilon$ for our new approach, we can still view this as an extremely good Nash equilibrium approximation. The difference is likely due to a combination of approximation error, roundoff error, and the fact that the GUI only displays decimals to four significant digits. The logit QRE solution differed significantly from the solution of our new approach (which is not too surprising since the game has infinitely many Nash equilibria). 
The logit solution gives expected payoff to player 1 of $\approx -0.0276$, player 2 of $\approx -0.0208$, and player 3 of $\approx 0.0484.$ The logit solution does fall in the infinite family from prior work subject to numerical precision (e.g., the logit solution has $a_{33} = 0.5002,$ while it is specified to be $\frac{1}{2}$ in the infinite family). When we ran the logit approach on the full version the GUI hung for 10 minutes, but the command line version output a solution after 2 minutes and 34 seconds. The command line output was difficult to comprehend and evaluate, but is presumably a close approximation of a Nash equilibrium. Thus, the logit QRE approach took 17.4 times longer to solve the reduced version than our new algorithm, but was successfully able to solve the full version in several minutes while our algorithm was not. This indicates that logit QRE is potentially more scalable and a suitable approximation algorithm for larger games, while our bilinear NCP approach is superior for exact Nash equilibrium computation in smaller games. 

\section{Conclusion}
\label{se:conc}
Computing Nash equilibrium in multiplayer imperfect-information games is perhaps the ``last frontier'' for solving standard game classes, as effective procedures have been devised for two-player strategic-form and extensive-form games and multiplayer strategic-form games. Many important scenarios contain more than two participants and imperfect information, so it is pivotal to have efficient and accurate methods to compute relevant solution concepts. 

We presented a new approach that solves a bilinear program based on a nonlinear complementarity problem formulation from the sequence-form game representation. The approach capitalizes on recent advances for solving nonconvex quadratic programs. Our algorithm is able to find a Nash equilibrium in three-player Kuhn poker after removal of dominated actions in under 3 seconds, while only one of the algorithms from Gambit's software suite was able to solve the game and it took over 40 seconds. 

As we have seen very recently with multiplayer strategic-form games, a new method based on solving a bilinear program~\cite{Ganzfried24:Fast} was quickly and dramatically improved upon by other bilinear approaches~\cite{Zhang23:Computing,Cerny25:Spatial}. It is quite likely that such improvements can also be made for extensive-form games. We would also expect that algorithms specialized for solving nonlinear complementarity problems should be able to outperform Gurobi's general nonconvex quadratic optimization method, though this is somewhat unclear based on the performances of standard Gambit approaches.

We do note that the scalability of our approach is limited; it is unable to solve the full version of three-player Kuhn poker which contains about twice as many nonzero entries in its optimization model, while the logit QRE approach was able to solve the full game in several minutes. For very large multiplayer imperfect-information games, the only option for Nash equilibrium approximation is to run an iterative self-play algorithm such as counterfactual regret minimization or fictitious play, which have no performance guarantees; however, once strategies have been computed it is easy to verify their degree of accuracy. Recent work has shown that for multiplayer strategic-form games using multiple initializations leads to a significant improvement in Nash equilibrium approximation of fictitious play~\cite{Ganzfried22:Fictitious}. It is very possible this would also hold for imperfect-information games and counterfactual regret minimization. If the game is small and we would like to guarantee finding an exact Nash equilibrium, then our results suggest that our new approach is the best option. For moderate-sized games that our algorithm is unable to solve, the best approach with a global convergence guarantee appears to be logit QRE tracing. 

\bibliographystyle{plain}
\bibliography{C://FromBackup/Research/refs/dairefs}

\end{document}